\documentclass[fleqn,usenatbib]{mnras}

\usepackage[T1]{fontenc}

\DeclareRobustCommand{\VAN}[3]{#2}
\let\VANthebibliography\thebibliography
\def\thebibliography{\DeclareRobustCommand{\VAN}[3]{##3}\VANthebibliography}

\usepackage{graphicx}	
\usepackage{amsmath}	
\usepackage{amssymb}	
\usepackage{adjustbox}  
\usepackage{newtxtext,newtxmath}

\title[LSS from non-Gaussian PDFs]{The large-scale structure from non-Gaussian primordial perturbations}

\author[G. A. Pe\~{n}a \& G. N. Candlish]{
G. A. Pe\~{n}a,$^{1}$\thanks{E-mail: greco.pena@postgrado.uv.cl}
G. N. Candlish,$^{1}$
\\
$^{1}$Instituto de F\'isica y Astronom\'ia, Universidad de Valpara\'iso, Gran Breta\~na 1111, Valpara\'iso, Chile\\
}

\date{Accepted XXX. Received YYY; in original form ZZZ}

\pubyear{2021}

\begin{document}
\label{firstpage}
\pagerange{\pageref{firstpage}--\pageref{lastpage}}
\maketitle

\begin{abstract}
The late-time effect of primordial non-Gaussianity offers a window into the physics of inflation and the very early Universe. In this work we study the consequences of a particular class of primordial non-Gaussianity that is fully characterized by initial density fluctuations drawn from a non-Gaussian probability density function, rather than by construction of a particular form for the primordial bispectrum. We numerically generate multiple realisations of cosmological structure and use the late-time matter power spectrum, bispectrum and trispectrum to determine the effect of these modified initial conditions. We show that the initial non-Gaussianity has only a small imprint on the first three polyspectra, when compared to a standard Gaussian cosmology. Furthermore, some of our models present an interesting scale-dependent deviation from the Gaussian case in the bispectrum and trispectrum, although the signal is at most at the percent level. The majority of our models are consistent with CMB constraints on the trispectrum, while the others are only marginally excluded. Finally, we discuss further possible extensions of our study.
\end{abstract}




\section{Introduction}

In the standard model of cosmology, inflation is assumed to generate the initial inhomogeneities that lead to the formation of structure in our Universe. It is, nonetheless, a considerable challenge to use cosmological observations to constrain the details of the inflationary model. Furthermore, from particle physics a wide variety of inflationary models have been postulated, thus motivating the need to find some means to discriminate between them. One approach that has received a great deal of attention is the study of primordial non-Gaussianity. While ``vanilla'' single scalar field inflationary models predict unobservably small deviations from Gaussianity, there are a number of generalisations to the standard inflationary scenario that lead to measureable levels of non-Gaussianity, such as multifield inflation, non-canonical kinetic terms and violations of slow-roll, among others (for comprehensive reviews of non-Gaussianity arising in inflationary models see \citealp{Bartolo2004,Chen2010,Celoria2018}).

A standard way of exploring these non-Gaussianities in the large-scale structure is through the polyspectra (the Fourier-space counterparts to the real-space $n$-point correlation functions). Many studies have explored the use of these statistics to attempt to constrain the primordial non-Gaussianity with late-time clustering observables. Most progress has been achieved in this area using cosmological N-body simulations (\citealp{Grossi2007,Grossi2009,Hikage2008,Desjacques2009,Giannantonio2010,Sefusatti2010,Wagner2010,Wagner2012} among others) as well as comparing with observations \citep{GilMarin2015,GilMarin2017}. While the extrapolation of analytical perturbation methods to the fully non-linear regime is problematic, multiple studies have explored this avenue, particularly with respect to the higher order correlation functions, such as \cite{Matarrese2007,Pietroni2008} and \cite{Assassi2015,Baldauf2015,Angulo2015} using the effective field theory of large-scale structures (EFT of LSS). Other studies have considered the full probability density function (PDF) of late-time structure to try to constrain the degree of primordial non-Gaussianity (\citealp{Valageas2002,Uhlemann2018,Friedrich2020}). 

The most common approach to including non-Gaussianity in cosmological simulations has been to determine the consequences of inflationary models that generate significant primordial non-Gaussianity for the Fourier space equivalent of the $3$-point correlation function, which is called the (primordial) bispectrum. The non-Gaussianity may be parameterised in the gravitational potential that is then used to generate the initial conditions for the N-body simulation via the usual approaches of Lagrangian perturbation theory \citep{scoccimarro2012,Celoria2018}.

In this work we will consider a different, and conceptually simpler, approach to test the effect of primordial non-Gaussianity on the formation of large-scale structure. We will use a recently-proposed characterisation of this primordial non-Gaussianity in terms of a non-Gaussian PDF describing the primordial curvature perturbations \citep{PDFpaper,Chen2018b}. The possible consequences for these models for large-scale structure has been studied \cite{Palma2020}. Using this approach will allow us to generate appropriate initial conditions for our simulations by simply changing the PDF from which we draw our initial sample of density perturbations. All other steps in the contruction of the initial conditions for our simulations (convolution with a transfer function and then application of perturbation theory to find the initial particle distribution) is then handled in the standard way. A significant difference of this approach compared to previous approaches to studying non-Gaussianity in cosmological simulations is that we do not limit ourselves to specifying a particular form of the primordial bispectrum, as we have the full PDF available to us.

Given that we wish to study the consequences of the primordial non-Gaussianity on structure formation, our numerical simulations involve only dark matter. In this work we aim to perform a preliminary exploration of the parameter space of this novel characterisation of the primordial non-Gaussianity, but in order to extract the signal from that induced through the non-linear process of gravitational collapse we must average over several realisations of cosmological evolutions. Thus we would, in principle, require multiple N-body simulations where we vary the initial sample of random numbers as well as the parameters of the PDF. To facilitate the generation of these realisations we have used the L-PICOLA\footnote{\href{https://github.com/CullanHowlett/l-picola}{https://github.com/CullanHowlett/l-picola}} code \citep{l-picola}, which has been verified to be sufficiently accurate for our purposes by comparing with a smaller number of full N-body simulations run using the RAMSES\footnote{\href{https://bitbucket.org/rteyssie/ramses}{https://bitbucket.org/rteyssie/ramses}} code \citep{ramses}.

This paper is organised as follows: in Section~\ref{sec:sims} we discuss the details of our simulations, specifically the method of generating the initial conditions, the way in which we explore the parameter space and the analysis we will perform on our results; in Section~\ref{sec:results} we present the details of that analysis; in Section~\ref{sec:conc} we discuss our results and future developments and give our conclusions.

\section{Simulations}
\label{sec:sims}

\subsection{Initial conditions}

\begin{figure*}
\centering
\includegraphics[width=1.\linewidth]{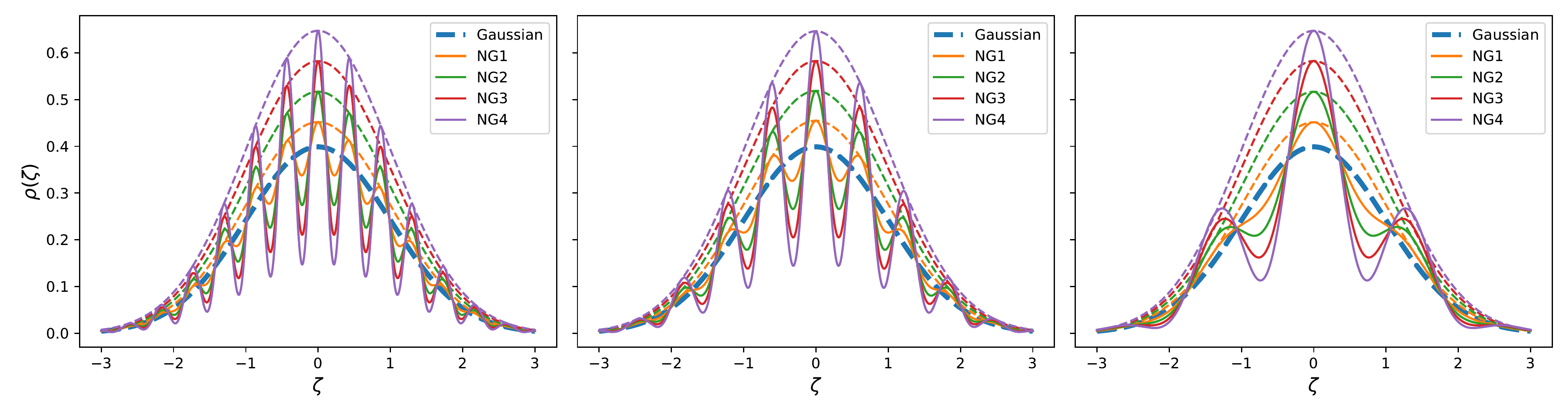}
\caption{The non-Gaussian probability density distributions for the primordial curvature perturbations considered in this work (equation~\ref{asymptotic pdf}). The thick blue dashed line is the fiducial Gaussian distribution. All of our PDFs have unit variance. The associated parameter values are given in Table~\ref{tab:A_f}.}
\label{PDF_plots}
\end{figure*}

The specific inflationary scenario we consider (introduced and thoroughly examined in \citealp{PDFpaper,Chen2018b}) comprises perturbations in an axion-like isocurvature field (with sinusoidal potential) which are coupled, through a linear derivative coupling, to the curvature field. Such models are physically very well motivated: axions have received much attention recently and arise naturally in UV-completions of gravity, such as string theory. The type of coupling considered in this model also arises naturally in a class of effective field theories that result from string compactifications.

This scenario gives rise to non-Gaussian initial conditions which are fully characterised by the probability density distribution of the primordial curvature perturbations. Assuming we can in principle access all scales in observations (i.e. $k \to \infty$) the PDF of the primordial curvature perturbations is determined to be
\begin{equation}\label{asymptotic pdf}
    \rho(\zeta)=\frac{e^{-\frac{\zeta^{2}}{2 \sigma_{\zeta}^{2}}}}{\sqrt{2 \pi} \sigma_{\zeta}}\left[1+A^{2} \frac{\sigma_{\zeta}^{2}}{f_{\zeta}^{2}} \cos \left(\frac{\zeta}{f_{\zeta}}\right)-A^{2} \frac{\zeta}{f_{\zeta}} \sin \left(\frac{\zeta}{f_{\zeta}}\right)\right].
\end{equation}
This form is referred to in \cite{PDFpaper} as the ``asymptotic form''. The full reconstructed PDF given in \cite{PDFpaper} considers a window function in $k$, i.e. a range of observationally accessible scales within an IR cutoff at long scales set by the size of the Universe, $k_{\text{IR}}$ and a small scale cutoff $k_{L}$ due to considering only the superhorizon perturbations during inflation. The full PDF has the form
\begin{equation}
\begin{split}
   \rho(\zeta)=\frac{e^{-\frac{\zeta^{2}}{2 \sigma_{\zeta}^{2}}}}{\sqrt{2 \pi} \sigma_{\zeta}}&\left[1+A^{2} \int_0^{\infty} \frac{dx}{x} \mathcal{K}_{\xi}(x) \right. \\
   &\left. \left( \frac{\sigma_{\zeta}^{2}}{f^2_{\zeta}(x)} \cos \left(\frac{\zeta}{f_{\zeta}(x)}\right)- \frac{\zeta}{f_{\zeta}(x)} \sin \left(\frac{\zeta}{f_{\zeta}(x)}\right)\right) \right]. 
\end{split}
\label{eq:fullPDF}
\end{equation}
where
\begin{equation}
\mathcal{K}_{\xi}(x) = \frac{2G(\xi,x)\ln \xi}{\pi F(\xi,x)}\exp \left( -\frac{\sigma_{\zeta}^2\left(f^2_{\zeta}(x)-f_{\zeta}^2\right)}{2f^2_{\zeta}f^2_{\zeta}(x)} \right),
\end{equation}
and
\begin{equation}
\begin{split}
G(\xi,x) &= \sin(x) - x \cos(x) - \sin(x/\xi) + (x/\xi) \cos (x/\xi) \\
F(\xi,x) &= \text{Ci}(x) - \frac{\sin(x)}{x} - \text{Ci}(x/\xi) + \frac{\sin(x/\xi)}{x/\xi}.
\end{split}
\end{equation}
$\text{Ci}(x)$ is the cosine integral function, and $\ln \xi \approx 8$ for scales accessible via the CMB (we assume this value for all subsequent analysis). For practical reasons it is much easier to generate our initial conditions using the asymptotic PDF of equation~(\ref{asymptotic pdf}) rather than equation~(\ref{eq:fullPDF}). Fortunately, however, the functional form of equation~(\ref{eq:fullPDF}) is extremely similar to that of equation~(\ref{asymptotic pdf}) when the parameter $A$ is appropriately rescaled. We will return to this later when we compare our models with CMB constraints in Section~\ref{sec:CMBconstraints}.

We thus now concentrate on the asymptotic form of the PDF, with $\sigma_{\zeta}^2 = 1$ always. Using this PDF, we can generate the initial conditions by drawing samples from these non-Gaussian distributions. The parameters we have chosen for the PDF given in equation~(\ref{asymptotic pdf}) are summarised in Table~\ref{tab:A_f}. The associated PDFs are shown in Fig~\ref{PDF_plots}. Note that we are considering large amplitude deviations from Gaussianity to explore the feasibility of detecting effects in the late-time large-scale structure. We will discuss observational constraints on these PDFs in Section~\ref{sec:CMBconstraints}.

\begin{table}
    \scalebox{0.83}{
    \begin{tabular}{l|l|l|l|l}
        Frequency & $A^2$, Level 1 & $A^2$, Level 2 & $A^2$, Level 3 & $A^2$, Level 4 \\
        \hline
        f1 ($f=7\times10^{-2}$) & $6.5\times10^{-4}$    & $1.45\times10^{-3}$       & $2.225\times10^{-3}$      & $3.05\times10^{-3}$  \\
        f2 ($f=1\times10^{-1}$) & $1.4\times10^{-3}$    & $3\times10^{-3}$          & $4.6\times10^{-3}$        & $ 6.2\times10^{-3}$   \\
        f3 ($f=2.4\times10^{-1}$) & $7.6\times10^{-3}$  & $1.7\times10^{-2}$        & $2.65\times10^{-2}$       & $3.585\times10^{-2}$    \\
    \end{tabular}}
    \caption{Values of $A^2$ and $f$ used in the non-Gaussian correction term of the PDF for our models.}
    \label{tab:A_f}
\end{table}

The thick blue dashed line indicates a Gaussian PDF with a standard deviation of unity. The orange, green, red and purple lines correspond to the four levels of non-Gaussianity we consider. We can clearly see that the modified PDF given in equation~(\ref{asymptotic pdf}) corresponds to a Gaussian that is modulated with an oscillatory component. This ensures that the odd-$n$ moments of the distribution are, in fact, equal to zero, as is the case for a pure Gaussian. The dashed lines correspond to Gaussian envelopes that touch the upper peaks of the oscillations. Modification of the parameter $f$ in equation~(\ref{asymptotic pdf}) results in a change to the frequency of the oscillatory modulation (and to the amplitude). We have thus altered the value of the parameter $A$ when the frequency is changed to ensure the central peak always takes the same value, for comparison purposes.

We generate a sample of each of our non-Gaussian PDFs by using a simple accept-reject technique, whereby we generate a uniformly distributed random value within a bounding box that includes the highest peak of the oscillatory PDF and is cut in the horizontal range between $\zeta_{\text{min}} = -4$, and $\zeta_{\text{max}} = 4$. After generating this value we simply compare with both the fiducial Gaussian distribution (thick dashed blue line in Fig.~\ref{PDF_plots}) as well as the non-Gaussian PDF. The value is then flagged according to whether it belongs to the Gaussian sample, to the non-Gaussian sample, to both, or to neither. In the latter case the value is rejected. This procedure is continued until we have generated $N$ values for both distributions, where $N$ is the total number of particles in our simulations, chosen to be equal to the number of points used in our discretised density field mesh. We choose $N = 256^3$ for all simulations in this work. Upon completion of the sample for one of the distributions we continue to add only points that belong to the other (e.g. if the Gaussian sample is completed, we reject any further points that belong to the Gaussian distribution and accept only those points that belong exclusively to the non-Gaussian distribution). Finally we order the non-Gaussian values according to the ordering of the Gaussian values. Specifically we sort both lists of numbers in ascending order, then reverse this sorting to recover the original ordering of the Gaussian values. The mapping required to perform this reversal is then applied to the non-Gaussian values. In this way the non-Gaussian values "inherit" the ordering of the Gaussian values. This helps to reduce the difference between the Gaussian and non-Gaussian samples, ensuring that, point-for-point in the density field, there is a minimal variation between the Gaussian and non-Gaussian simulations. This then produces initial conditions that subsequently lead to very similar evolution due to gravitational instability, forming closely comparable structures at all redshift, with the small differences primarily arising from the effect of the non-Gaussianity. Furthermore, we will attempt to mitigate the gravitational contribution to the non-Gaussianity arising from non-linear structure formation, which hides the signal coming from the primordial non-Gaussianity, by normalizing our statistics with the Gaussian model, as we will discuss later.

The procedure outlined above is then repeated for each non-Gaussian PDF that we consider. We use $4$ different values of $A$ (which we refer to as \textit{levels} of non-Gaussianity), $3$ different values of $f$ and $5$ different initial seeds for the uniform random number generator. This leads to $60$ non-Gaussian models, as well as $5$ fiducial Gaussian models, one for each choice of seed for the random number generator. We refer to each of the simulations using a different seed as a \textit{realisation}.

To generate the initial conditions for RAMSES these values are then passed to the MUSIC\footnote{\href{https://bitbucket.org/ohahn/music}{https://bitbucket.org/ohahn/music}} code, where they are Fourier transformed and convolved with a transfer function generated by CAMB\footnote{\href{https://github.com/cmbant/CAMB}{https://github.com/cmbant/CAMB}} \citep{camb}. We thus calculate the convolution
\begin{equation}
\delta(\vec{k}) = c k^{n_s/2} T(k) \mu(\vec{k}),
\end{equation}
in the MUSIC code, where $\mu(\vec{k})$ is the Fourier-transformed set of random numbers. Note that it is standard practice in cosmological N-body simulations to use a unit variance for the white noise field $\mu(\vec{k})$, given that the normalisation and variance of the density field $\delta(\vec{k})$ are constructed to correspond to a given matter power spectrum. For this reason we have set $\sigma_{\zeta} = 1$ for all our primordial PDFs. The transfer function $T(k)$ corresponds to a standard LCDM cosmology, and $c$ is a normlisation constant. For the runs using L-PICOLA we have modified the initial conditions generation of that code, which, as in MUSIC, uses the 2LPT method \citep{2LPT}. The modifications allow us to read in the same random number samples as used for MUSIC and use these in the convolution operation (also performed in Fourier space) so that the process of initialising the L-PICOLA simulation is exactly the same as that performed by MUSIC.

To identify easily the models that we are considering in our analysis, we have adopted a short-hand naming convention to indicate the properties of each model:
\begin{itemize}
    \item Initial condition type: \textbf{G} for Gaussian, \textbf{NG} for non-Gaussian.
    \item Level of non-Gaussianity: 1, 2, 3 or 4. This only applies for NG models.
    \item Frequency of non-Gaussianity: f1, f2 or f3. Again this only applies for NG models.
    \item Random realisation: r1, r2, r3, r4 or r5. This will only be necessary when we refer to one specific realisation. For most results we average over all realisations.
\end{itemize}
Thus, for example, the second realisation of the non-Gaussian model with $A=3 \times 10^{-3}$ and $f=0.1$ (corresponding to a level 2 model using the f2 frequency) would be referred to as NG2f2r2. We will typically not need to consider the specific realisation used.

\subsection{Realisations}

To statistically explore the effect of the non-Gaussianity we must generate multiple realisations of our cosmological volume. As this is computationally expensive we have opted to generate a small number of models using the RAMSES cosmological N-body code, with the rest of our models being generated using the mock catalogue generation code L-PICOLA. The models we have generated are summarised in Table 1. We use a standard LCDM cosmology with the following parameters in both RAMSES and L-PICOLA: $\Omega_{m} = 0.3$, $\Omega_{b} = 0.04$, $\Omega_{\Lambda} = 0.7$, $\sigma_{8} = 0.88$, and $n_{s} = 0.96$. In all cases we use a box size of $500$ Mpc and a particle resolution of $256^3$. The coarsest grid resolution in RAMSES is set to $256^3$ points, with $6$ further levels of refinement possible, using the AMR capabilities of the code, leading to a maximum force resolution of $\sim 30$ kpc. The (fixed) grid resolution used in L-PICOLA to calculate the density field is also set to $256^3$ points.

For the RAMSES simulations we have generated one realisation of the Gaussian model and the $4$ non-Gaussian models with frequency parameter fixed at f1. We also ran these same $5$ models (i.e. using precisely the same initial conditions) using our modified L-PICOLA code, for comparison purposes. The results of this comparison will be given in the next section. We have also run the same $4$ levels of non-Gaussianity for each frequency considered and each realisation (i.e. random number seed), plus a Gaussian run for each realisation. This results in a total of $65$ models generated in L-PICOLA: ($4$ levels of non-Gaussianity, $3$ different frequencies, $5$ different realisations, plus $5$ different Gaussian runs). In RAMSES we have one frequency (f1), one realisation (r1) and $5$ levels of non-Gaussianity (including zero), resulting in $5$ models. 

The polyspectra that we consider are all calculated using a modified version of the publicly-available Pylians\footnote{\href{https://github.com/franciscovillaescusa/Pylians3}{https://github.com/franciscovillaescusa/Pylians3}} code \citep{pylians}, where the calculation of the trispectrum has been added. Note that all of these polyspectra are estimated using the method of \cite{Watkinson}.

\subsection{Analysis}

We will concentrate our analysis on the lowest n-point statistics. Although the nature of the non-Gaussianity considered in this work is such that it cannot be fully characterised by these measures, they are nevertheless a useful tool used in many theoretical and observational studies of non-Gaussianity. It is thus important to determine the effect of our primordial NG model on the late-time $n$-point statistics, to see to what extent we can detect this primordial non-Gaussianity in these quantities. We leave for future work the study of other measures of non-Gaussianity that may be more sensitive to the primordial signal. It should be noted that the non-Gaussian PDF is constructed such that the odd-$n$ correlation functions are identically zero, just as in the case of a pure Gaussian PDF. The even-$n$ correlation functions contain corrections arising from the non-Gaussianity, and are thus no longer expressible purely in terms of the $2$-point function, as is the case for a Gaussian PDF. In the case of large-scale structure statistics, gravitational collapse induces non-Gaussianities which arise in all the n-point statistics. We are therefore interested in the possible signature of the primordial non-Gaussianity that we are considering in the power spectrum ($2$-point correlation function), the bispectrum ($3$-point correlation function) and the trispectrum ($4$-point correlation function).

While the power spectrum may be considered as a function of $k$ alone (due to isotropy), the bispectrum and trispectrum may contain, in principle, many more datapoints by considering all possible triangular configurations of $\mathbf{k}_1$, $\mathbf{k}_2$ and $\mathbf{k}_3$ (bispectrum) or quadrilateral configurations of $\mathbf{k}_1$, $\mathbf{k}_2$, $\mathbf{k}_3$ and $\mathbf{k}_4$ (trispectrum). In this work we limit ourselves to the equilateral case for the bispectra, where $k_1=k_2=k_3$ and what we refer to as the ``square'' case for the trispectra, where $k_1 = k_2 = k_3 = k_4$. A quadrilateral in $3$-dimensional space need not have all four vertices co-planar. There are thus two additional degrees of freedom, beyond the magnitudes $k_i$, to specify the trispectrum $k$-space configuration, which may be taken as the lengths of the diagonals connecting opposite vertices, given by $|\mathbf{k}_1+\mathbf{k}_2|$ and $|\mathbf{k}_2+\mathbf{k}_3|$. In other words, a $3$-dimensional quadrilateral may be folded. Our ``square'' configurations are only restricted in the side lengths. Thus, more precisely, we consider all folded quadrilaterals with equal side lengths. This is straightforward to implement in Pylians thanks to the method of \cite{Watkinson}.

For all polyspectra we consider $35$ linearly-spaced bins in the range $2.2k_F \leq k \leq k_F N_k/3$ where $N_k$ is the number of points used in our discretised Fourier space in each dimension, which we set to be equal to the particle resolution in our simulations, $N_{k} = 256$. The fundamental frequency accessible in our simulations is $k_F = 2\pi/L$, where $L$ is the box length. The factor of $1/3$ in our upper limit on $k$ is to ensure we do not consider very small scales where the estimator is expected to perform poorly \citep{Sefusatti2016}. This also ensures we are well below the Nyquist frequency of $k_{\text{Nyq}} = k_F N_k/2$. At the low $k$ end we try to avoid values of $k$ that are too heavily affected by sample variance, due to the small number of $k$-configurations (triangles for the bispectrum, squares for the trispectrum) by considering values somewhat larger than the fundamental frequency $k_F$.

Note that for the bispectra, throughout our analysis, we have removed the first value of $k$ in our range due to very large sample variance.

\section{Results}
\label{sec:results}

\subsection{Comparing RAMSES and L-PICOLA}

\begin{figure*}
\centering
\includegraphics[width=0.8\linewidth]{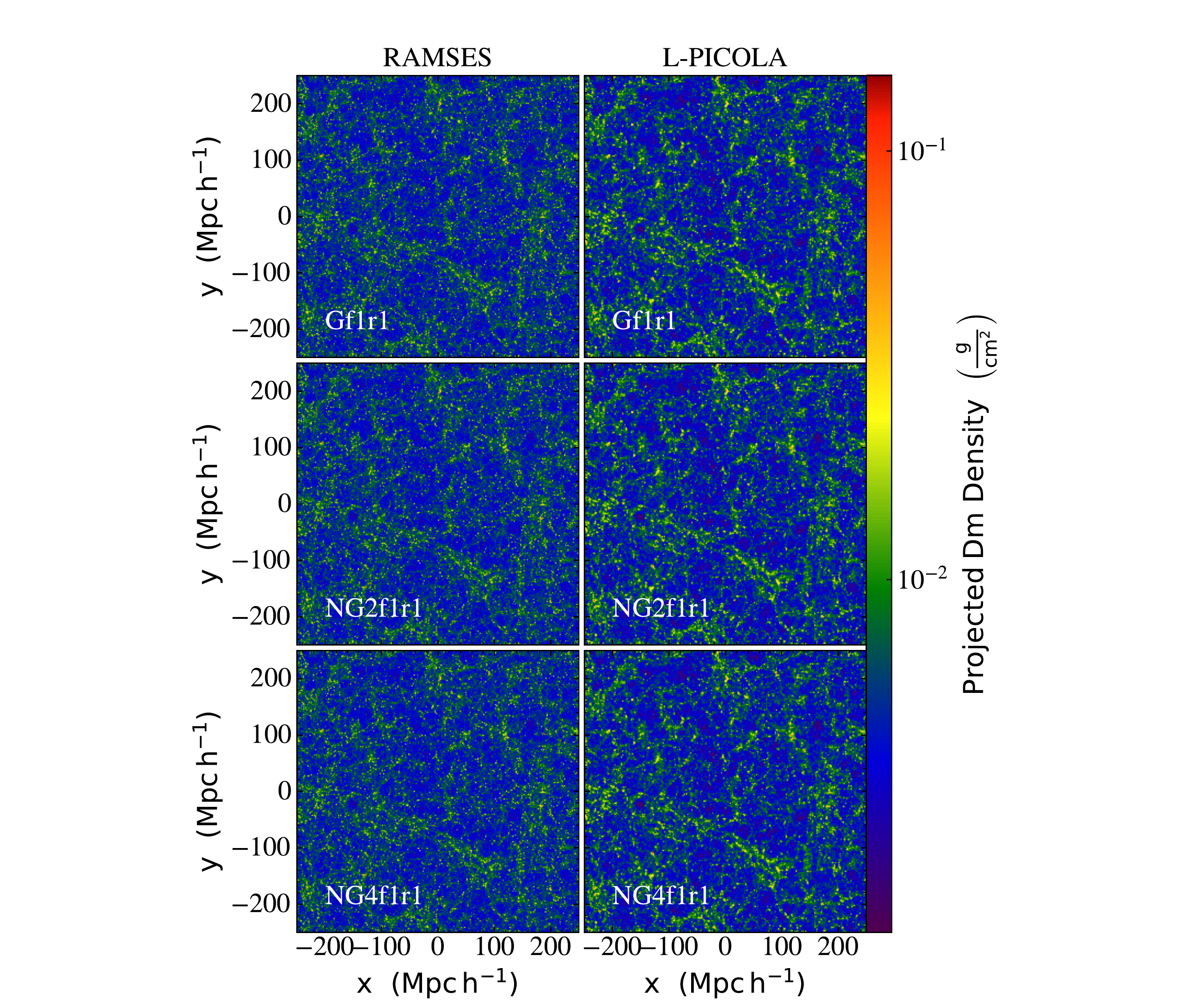}
\caption{Projected dark matter density for the RAMSES and L-PICOLA simulations at z=0. \textit{First row}: Gaussian initial conditions (model Gr1); \textit{second row}: the NG2f1r1 model; \textit{third row}: the NG4f1r1 model.}
\label{ram_pic}
\end{figure*}

In order to justify the use of the L-PICOLA code in this work, we will now compare the results with RAMSES by a visual comparison of the projected dark matter density distributions and by comparing the low-$n$ polyspectra. The projected density fields for RAMSES and L-PICOLA for one realisation are shown in Figure~\ref{ram_pic} at z=0. The left column shows the results using RAMSES while the right column shows the results using L-PICOLA. The first row is for the fully Gaussian initial conditions (model Gr1), the second row for the NG2f1r1 model and the last row for the NG4f1r1 model.

We can see that the overall distribution of structure is extremely similar for both the RAMSES and L-PICOLA runs, especially at larger scales. It is also worth noting that the non-Gaussianity does not lead to any visually obvious change in the density distribution. The small scale structures (halos) in all the simulations are clearly much more defined in RAMSES compared to L-PICOLA, as would be expected due to the full N-body calculation undertaken by the former code, which includes adaptive mesh refinement to improve the spatial resolution in the overdensities by more than an order of magnitude. The L-PICOLA calculation, in contrast, is based on a mesh with a fixed resolution, equivalent to the coarse level grid resolution used in our RAMSES runs.

\begin{figure*}
    \centering
    \includegraphics[width=0.8\linewidth]{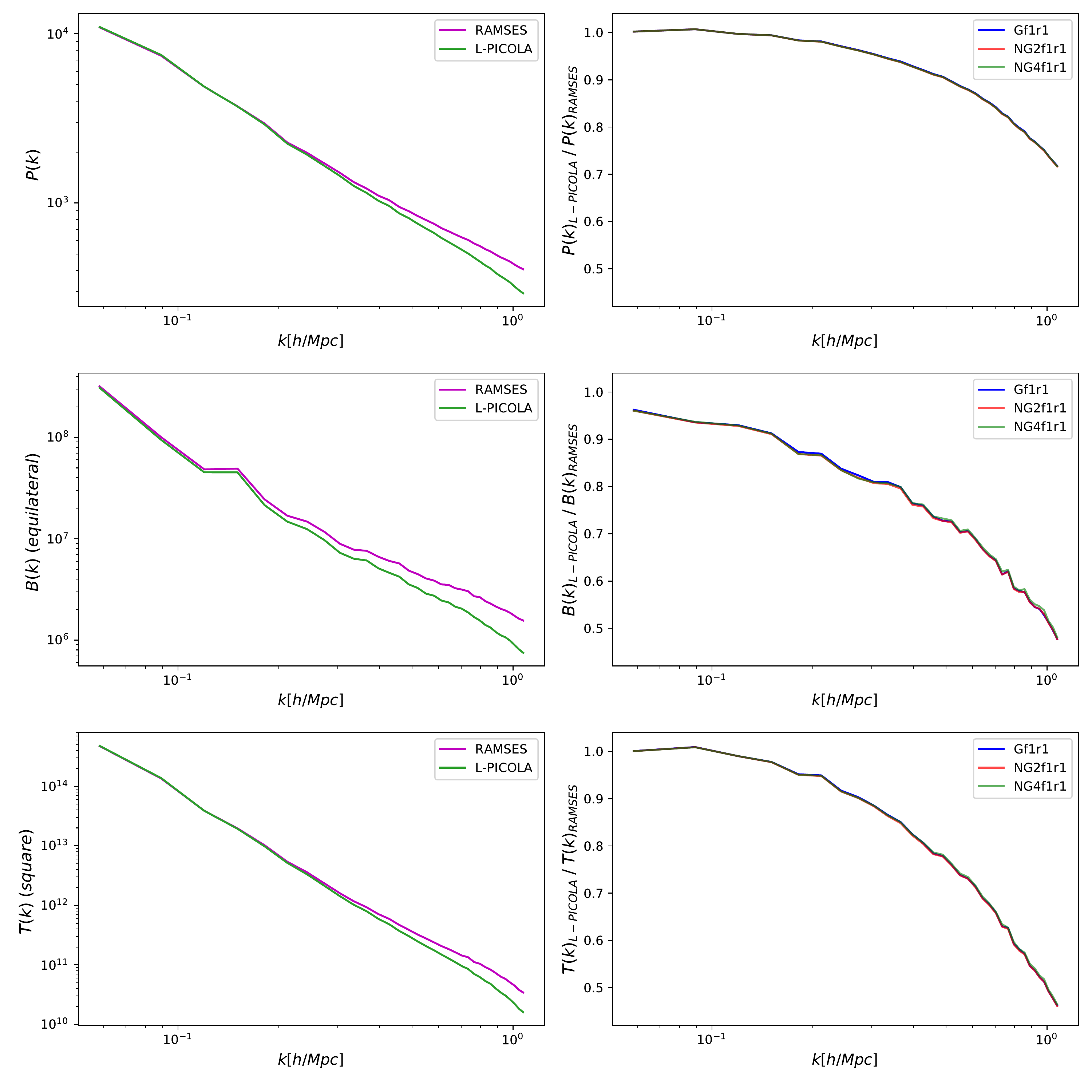}
    \caption{\textit{Left column}: polyspectra of a single realisation of a Gaussian simulation using RAMSES and L-PICOLA. \textit{Right column}: The polyspectra of L-PICOLA (for one realisation) normalised by those of RAMSES for models Gr1, NG2f1r1 and NG4f1r1. \textit{Top row}: power spectrum, \textit{middle row}: bispectrum, \textit{bottom row}: trispectrum. Only equilateral triangle configurations are considered for the bispectrum and only square configurations for the trispectrum.}
    \label{spectras_ram_pic}
\end{figure*}

The differences can be more clearly seen in the polyspectra calculated from the two codes, as shown in the left column of Fig.~\ref{spectras_ram_pic}. The power spectrum, bispectrum and trispectrum of a single realisation of a Gaussian model are shown, comparing the results using RAMSES and L-PICOLA. We can see that, for large scales (small $k$) there is good agreement between the codes, while for small scales (large $k$) there is some notable divergence. In the right column we show the L-PICOLA polyspectra for the Gr1, NG2f1r1 and NG4f1r1 models normalised by the polyspectra for the same models as derived from RAMSES. At large scales ($k < 0.2$) there is excellent agreement between the two codes, with differences for the power spectrum and trispectrum well within $5\%$, while the bispectrum shows somewhat more than $10\%$ difference. At the smallest scales ($k \gg 0$) the differences are on the order of $25\%$ for the power spectrum and $50\%$ for the bi- and trispectrum. An important result for our study is that the inclusion of non-Gaussianity does not affect these results, i.e. there is effectively no spread in the lines in the plots in the right column of Fig.~\ref{spectras_ram_pic}.

\begin{figure*}
    \centering
    \includegraphics[width=0.8\linewidth]{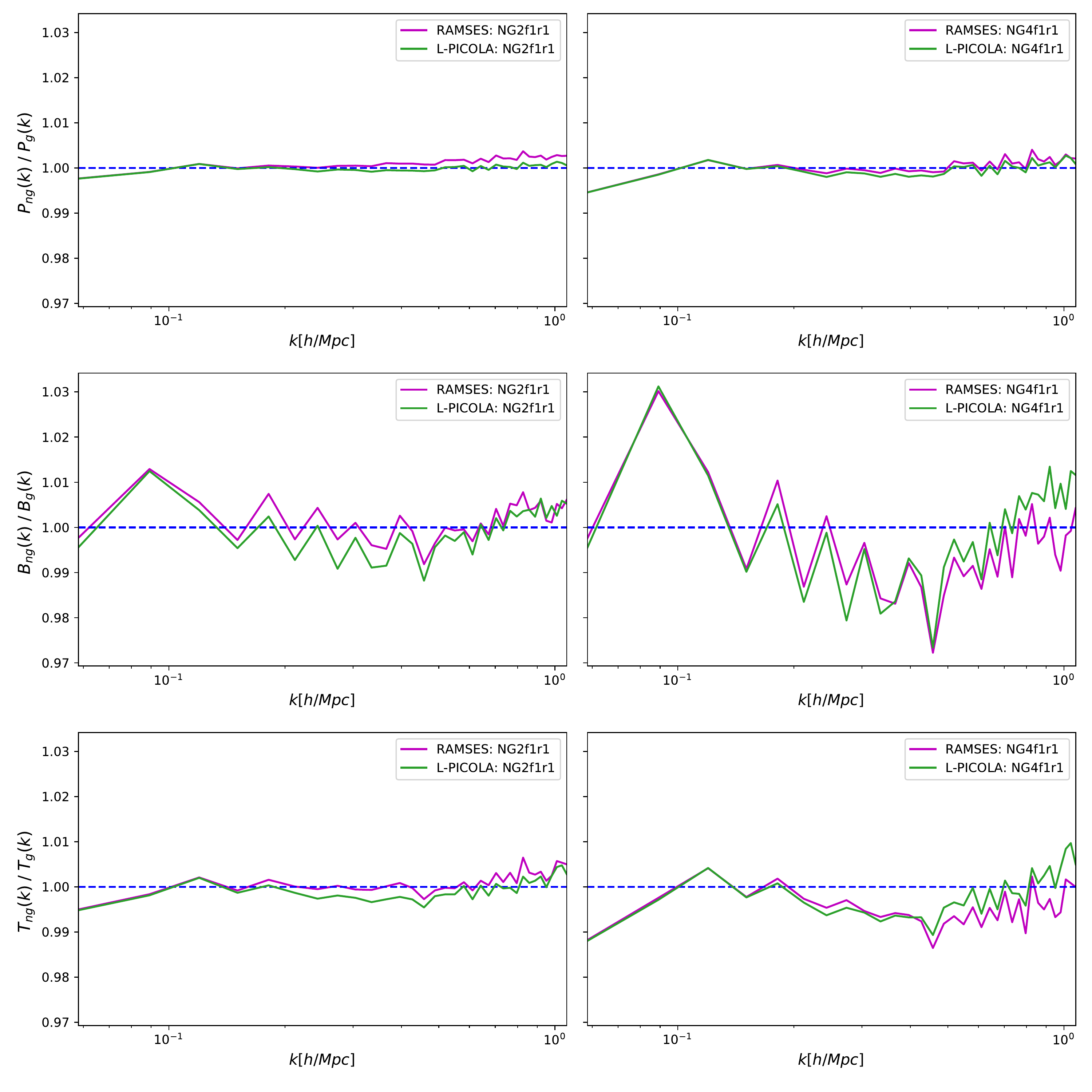}
    \caption{Comparison of the normalised polyspectra for L-PICOLA and RAMSES. \textit{Left column}: NG2f1r1; \textit{Right column}: NG4f1r1.}
    \label{spectras_ram_pic:levels}
\end{figure*}

In Fig.~\ref{spectras_ram_pic:levels} we show the polyspectra for the NG2f1r1 (left column) and NG4f1r1 models (right column) normalised with respect to the Gaussian models. We do this for both RAMSES and L-PICOLA. The advantage of working with these normalised quantities is that the discrepancy between the L-PICOLA results and those of RAMSES shown in Fig.~\ref{spectras_ram_pic} are effectively cancelled out as the large disagreement between these codes at small scales is independent of the level of non-Gaussianity, as discussed ealier. This is to be expected, as the lack of small scale structure is a general problem with L-PICOLA, regardless of the nature of the initial conditions. Indeed, in Fig.~\ref{spectras_ram_pic:levels} we can see that these normalised polyspectra agree extremely well across all scales, with differences well within $1\%$. This gives us high confidence that the non-Gaussian polyspectra normalised by the Gaussian polyspectra are very well represented by the L-PICOLA code. Throughout the rest of this paper we will refer to these quantities as \textit{normalised polyspectra}.

\subsection{Variance due to differing realisations}
\label{sec:variance}

To remove the variance arising from differing realisations of the large-scale structure, we determine an average normalised polyspectra by averaging over all $5$ realisations for each model. In subsequent plots, therefore, we will consider
\begin{equation}
 \Delta_{\mathcal{P}}(k) = \frac{1}{n} \sum_{i=1}^n \frac{\mathcal{P}^{(i)}_{ng}(k)}{\mathcal{P}^{(i)}_g(k)},
\end{equation}
where $\Delta_{\mathcal{P}}$ corresponds to an \textit{averaged normalised polyspectrum}, showing deviations from the Gaussian case, averaged over all realisations. $\mathcal{P}_g(k)$ is a Gaussian polyspectrum and $\mathcal{P}_{ng}(k)$ a non-Gaussian polyspectrum. As discussed earlier, we consider only symmetrical configurations of the wavenumbers in this work, so we may treat all polyspectra as depending upon a single value of $k$. The index $i$ refers to a realisation (i.e. a different initial seed for the random number generation), and $n=5$ is the total number of realisations.

In this way we can attempt to cancel out the contribution to these polyspectra arising from the non-linear structure formation, which will vary in each realisation. We then use the various realisations to estimate minimum and maximum values for the normalised polyspectra, which we will refer to as the variance around our averaged normalised polyspectra.

\subsubsection{Power spectra}
 
 \begin{figure*}
    \centering
    \includegraphics[width=0.8\linewidth]{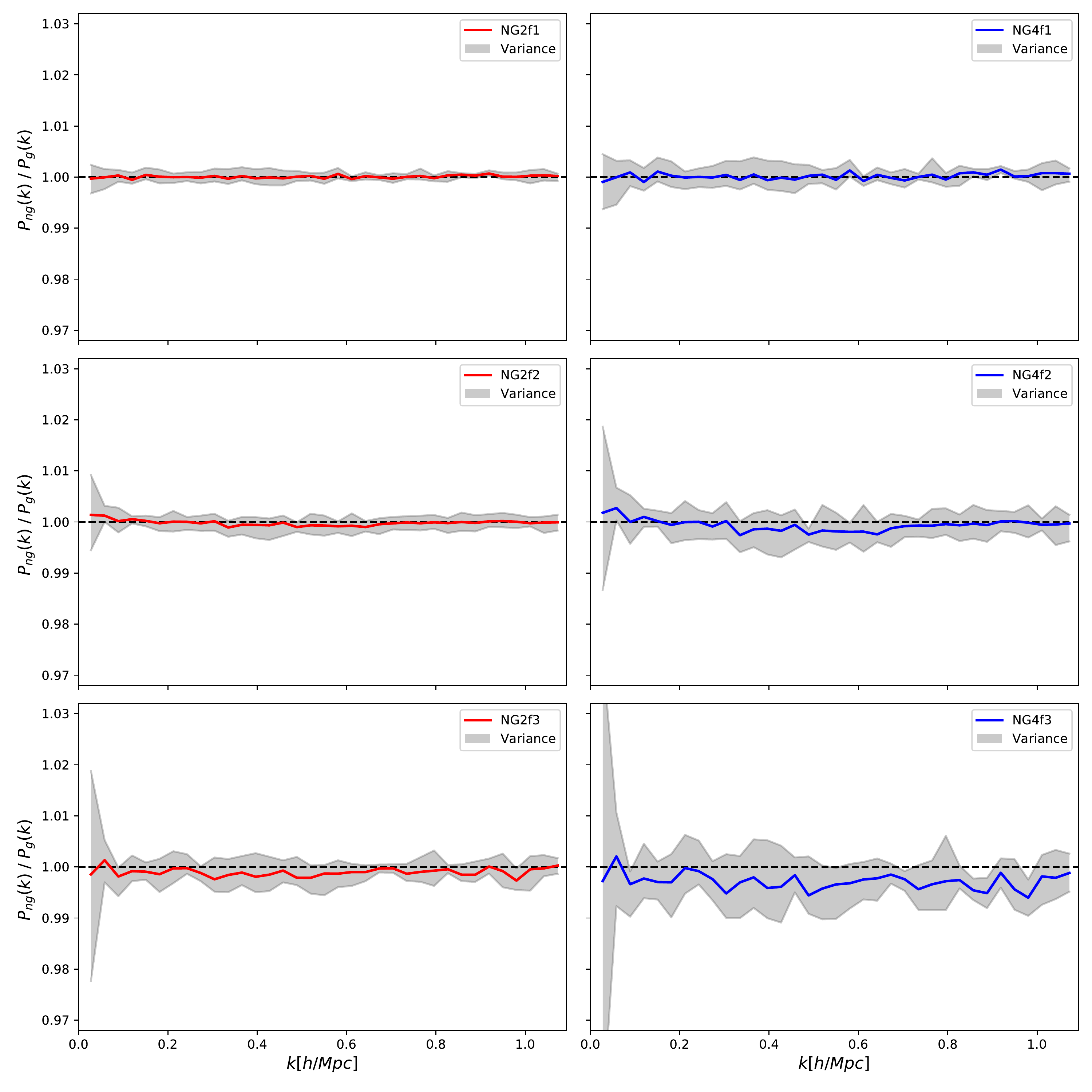}
    \caption{Non-Gaussian power spectra normalised with respect to the Gaussian power spectra, averaged across all realisations at $z=0$. \textit{Left column}: NG2 models, \textit{right column}: NG4 models. \textit{Top row}: frequency f1, \textit{middle row}: frequency f2, \textit{bottom row}: frequency f3. The red and blue lines indicate the average normalised power spectra (for the NG2 and NG4 models respectively) while the light grey strip shows the degree of variance around this average arising from the individual realisations.}
    \label{power_spec_realisations_z0}
\end{figure*}

\begin{figure*}
    \centering
    \includegraphics[width=0.8\linewidth]{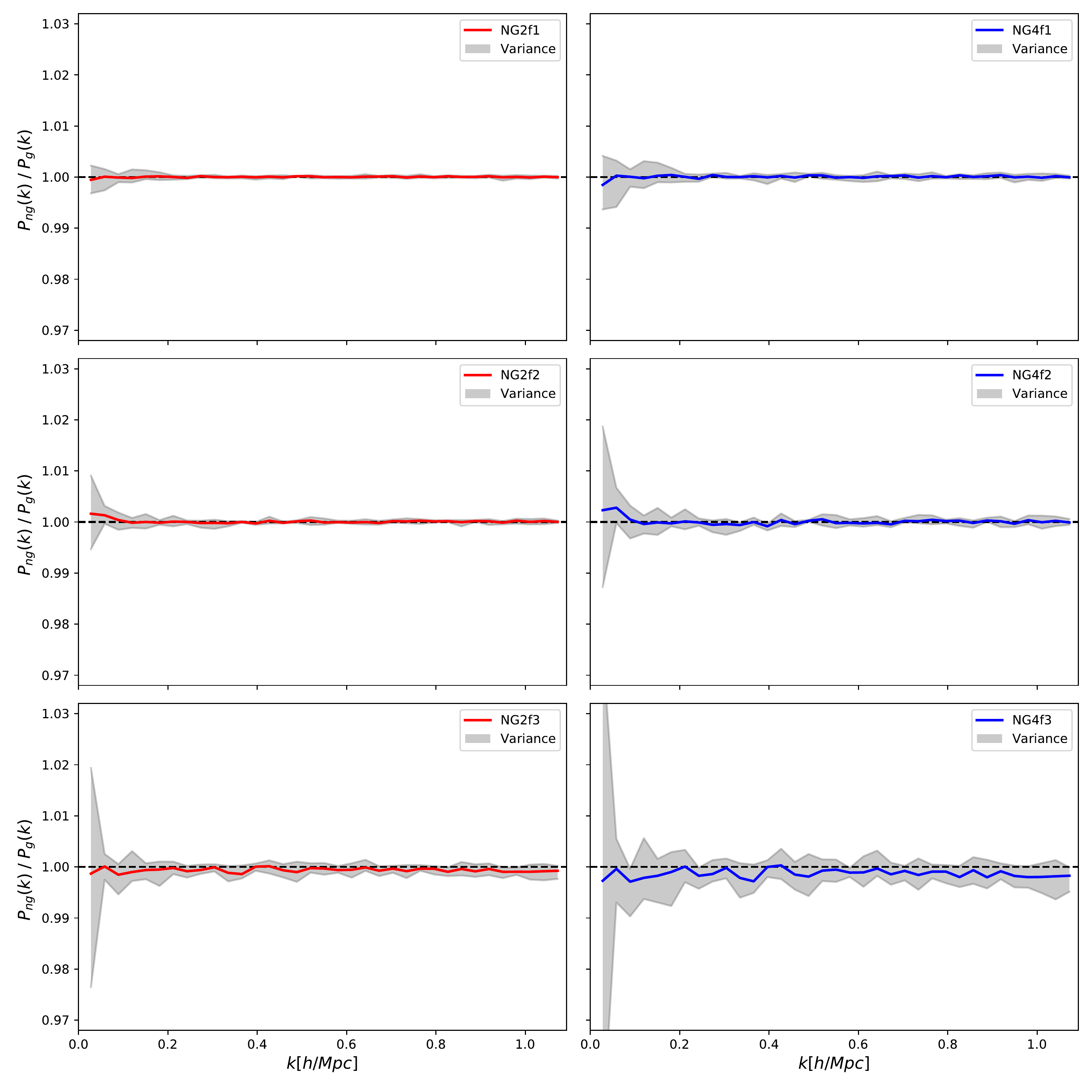}
    \caption{Non-Gaussian power spectra normalised with respect to the Gaussian power spectra, as for Fig.~\ref{power_spec_realisations_z0}, but at $z=2.2$.}
    \label{power_spec_realisations_z2.2}
\end{figure*}
 
In Fig. \ref{power_spec_realisations_z0} we show the averaged normalised power spectra, at $z=0$, for a selection of models. In the left column we show the results for the NG2 models, for frequency f1 in the top row, f2 in the middle row and f3 in the bottom row. Similarly, the results for these three frequencies for the NG4 models are shown in the right column. The red and blue lines indicate the average values across all realisations for the NG2 and NG4 models respectively, while the grey strips show the variance amongst realisations.

We can see that there is no scale dependence in these results, except for a very weak dependence in the NG4f2 model, which presents a feature that we refer to as a ``dip'' between $k \sim 0.2$ h/Mpc and $k \sim 0.8$ h/Mpc. Furthermore, the variance across realisations is always less than $1\%$, with the largest variance being manifest in the model with the strongest deviation from the Gaussian case (the NG4f3 model). There are some hints of a suppression in the power spectra, relative to the Gaussian case, in particular for the NG4f3 model, and this appears to persist across all scales that are probed by our simulations. Considering the PDFs being used for the NG4f2 and NG4f3 models, it appears that the shorter wavelength deviations from Gaussianity in the NG4f2 model lead to a (very weak) suppression of structure in a limited range of scales, whereas the long wavelength deviation exhibited in the PDF for the NG4f3 model leads to a suppression of power across a wider range of scales (at least within the limits of that which can be explored in our simulations). Note, however, that this suppression still lies within the variance band determined by considering all $5$ realisations. Thus this signal is at the boundary of what we could consider to be statistically significant. Nevertheless, it is noteworthy that such a large deviation from Gaussian statistics in the initial conditions (as exemplified by the NG4f3 model) fails to give rise to a significant deviation in the late-time matter power spectrum.

In Fig. \ref{power_spec_realisations_z2.2} we show a similar plot, but this time considering results for the power spectra at a higher redshift of $z=2.2$. Firstly we see that the variance is substantially diminished in all cases, consistent with the idea that this variance is primarily driven by the process of structure formation, and thus increases over time. We also see that there is essentially no significant deviation from equality with the Gaussian results.

\subsubsection{Bispectra}

\begin{figure*}
    \centering
    \includegraphics[width=0.8\linewidth]{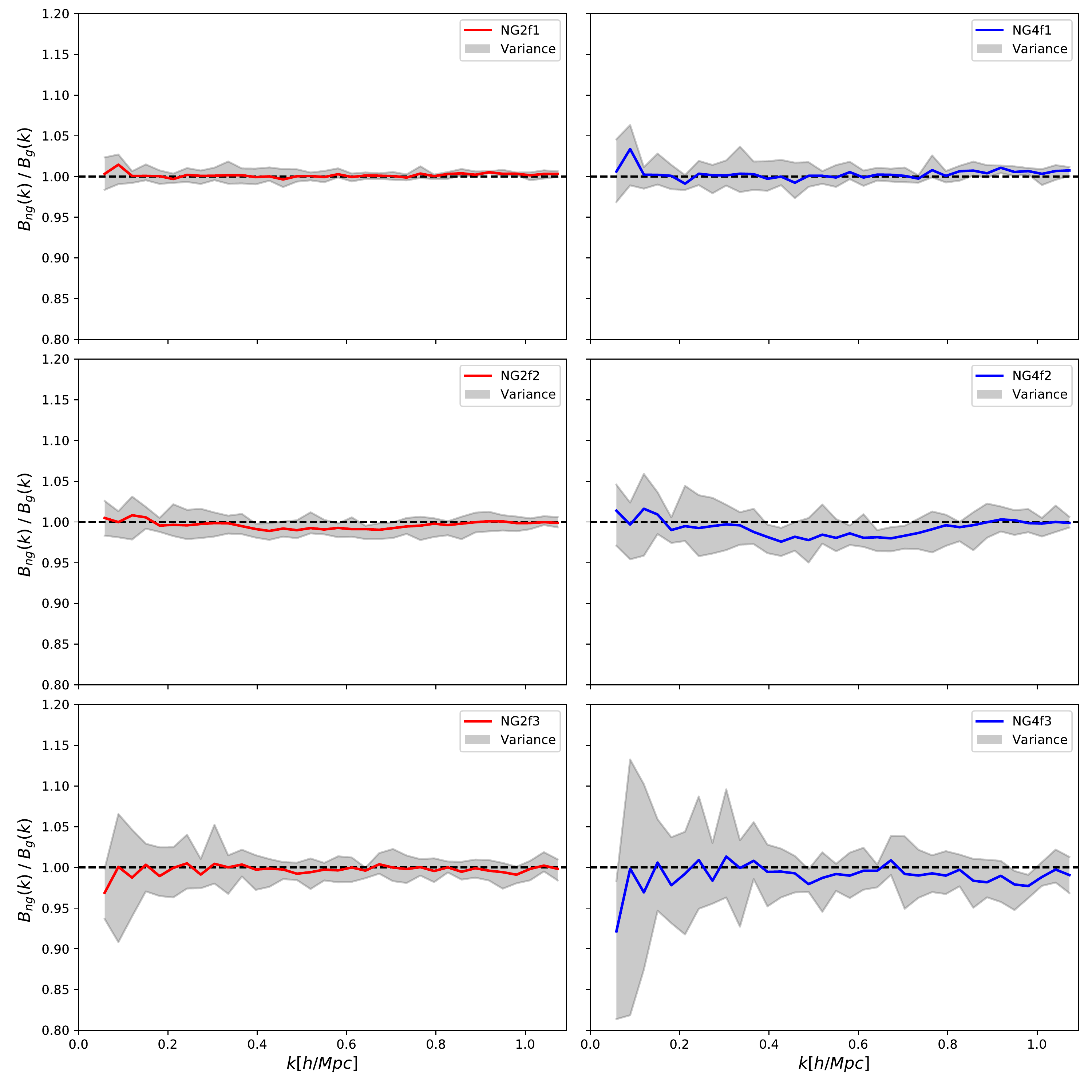}
    \caption{Non-Gaussian bispectra normalised with respect to the Gaussian bispectra, at $z=0$. The panels are as for Fig.~\ref{power_spec_realisations_z0}.}
    \label{bk_realisations_z0}
\end{figure*}

\begin{figure*}
    \centering
    \includegraphics[width=0.8\linewidth]{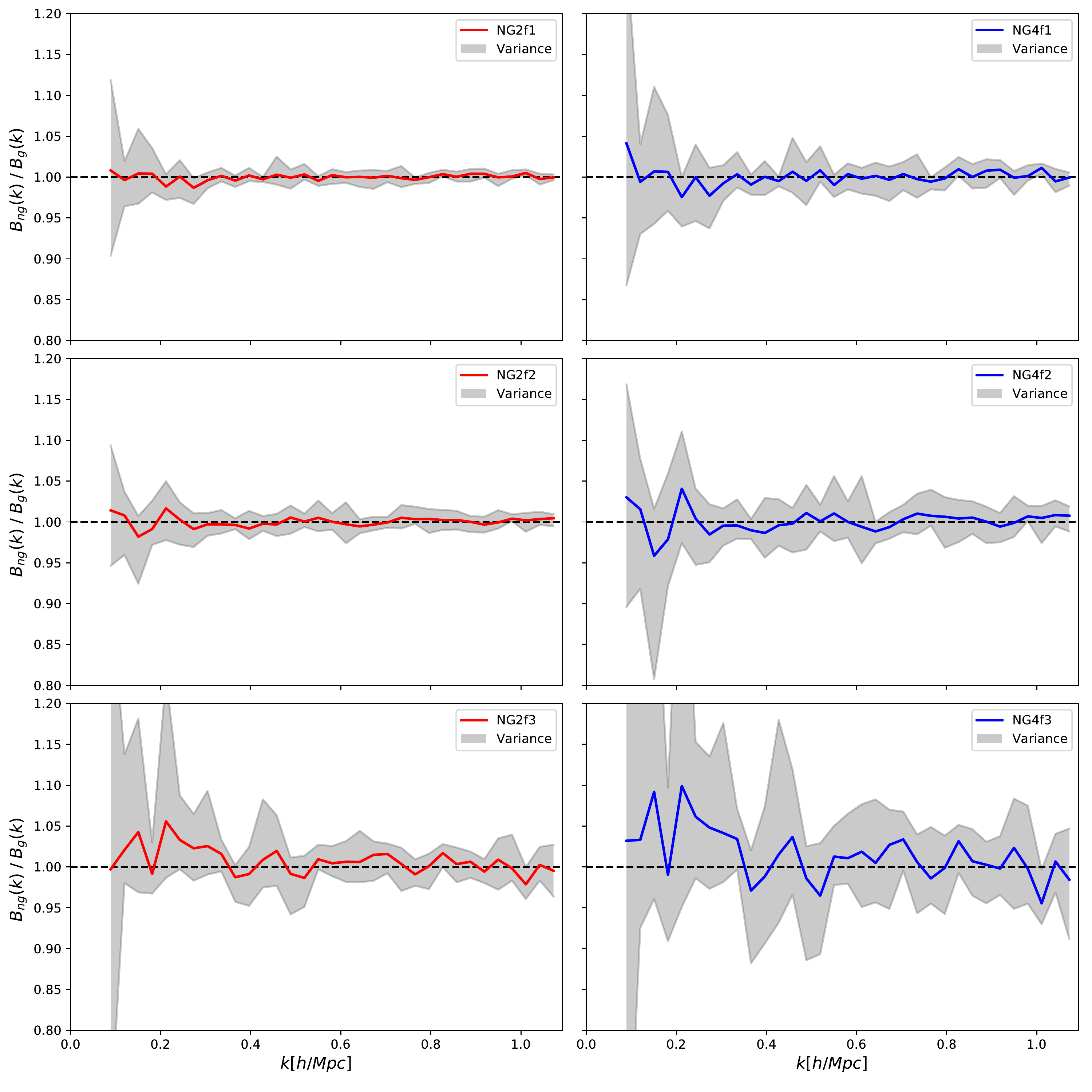}
    \caption{Non-Gaussian bispectra normalised with respect to the Gaussian bispectra, at $z=2.2$. The panels are as for Fig.~\ref{power_spec_realisations_z0}.}
    \label{bk_realisations_z2.2}
\end{figure*}

We now move on the next polyspectrum, the bispectrum. This generally depends on three wavenumbers $k_1$, $k_2$ and $k_3$. Unlike in the case of the power spectrum, we cannot appeal to isotropy to reduce the dependence to just one $k$. To simplify our analysis, however, we have decided to focus only on equilateral triangle configurations, where $k_1=k_2=k_3$. This allows us to treat the bispectrum as a function of a single value of $k$. The results for the averaged normalised bispectra, at $z=0$, are given in Fig.~\ref{bk_realisations_z0}. Again, there is very little scale dependence in these results, although it is interesting to note that the ``dip'' present at $0.2 \lesssim k \lesssim 0.8$ h/Mpc that is present for the frequency f2 models in the power spectrum (see Fig.~\ref{power_spec_realisations_z0}) is also present for the bispectrum, with a larger deviation from the equality line (black dashed line in the figure). It is also worth noting that the variance in these results across realisations is somewhat larger, particularly at larger scales. In the case of NG4f3 it is as large as $\sim 10\%$, whereas in all other models the bispectrum variance is mostly below $\sim 5\%$. This contrasts with the power spectra where the variance was always well below $1\%$.

Apart from the presence of the ``dip'' in the models using frequency f2, there is little evidence of deviation from the Gaussian case in the other models. The suppression of the power spectrum seen for NG4f3 is not repeated for the bispectrum.

It is important to take into account that the vertical axis scale in Fig.~\ref{bk_realisations_z0} is five times larger than that of Fig.~\ref{power_spec_realisations_z0}, thus the ``dip'' feature is more pronounced for the bispectrum than for the power spectrum. We will examine this more carefully in Section~\ref{sec:fitting_functions}. This leads to an interesting conclusion regarding the sensitivity of the bispectrum to this form of non-Gaussianity, which is by construction a symmetric oscillatory correction to an underlying Gaussian. This does not violate the property of Gaussian distributions that the odd-$n$ moments are identically zero. Thus the $3$-point correlation function (whose $k$-space analogue is the bisepctrum under consideration here) in perfect Gaussian conditions would vanish. Normally the presence of non-linear structure induces non-Gaussianities such that the bispectrum is not zero. What we apparently find, however, is that structure formation driven from this primordial non-Gaussianity imprints a marginally statistically significant signal in the bispectrum at the limit of detectability above the ``noise'' resulting from non-linearities, at least for this choice of parameters.

In Fig.~\ref{bk_realisations_z2.2} we consider the bispectra at $z=2.2$. Again we see no clear indication of any statistically significant deviation from the Gaussian case, although there is considerably more variance in this statistic at this redshift than seen for the power spectrum, rising to as much as $\sim 20\%$ for the NG4f3 model. This suggests that the bispectrum is particularly sensitive to the variations in the initial conditions, but that there is no specific systematic effect.

\subsubsection{Trispectra}

\begin{figure*}
    \centering
    \includegraphics[width=0.8\linewidth]{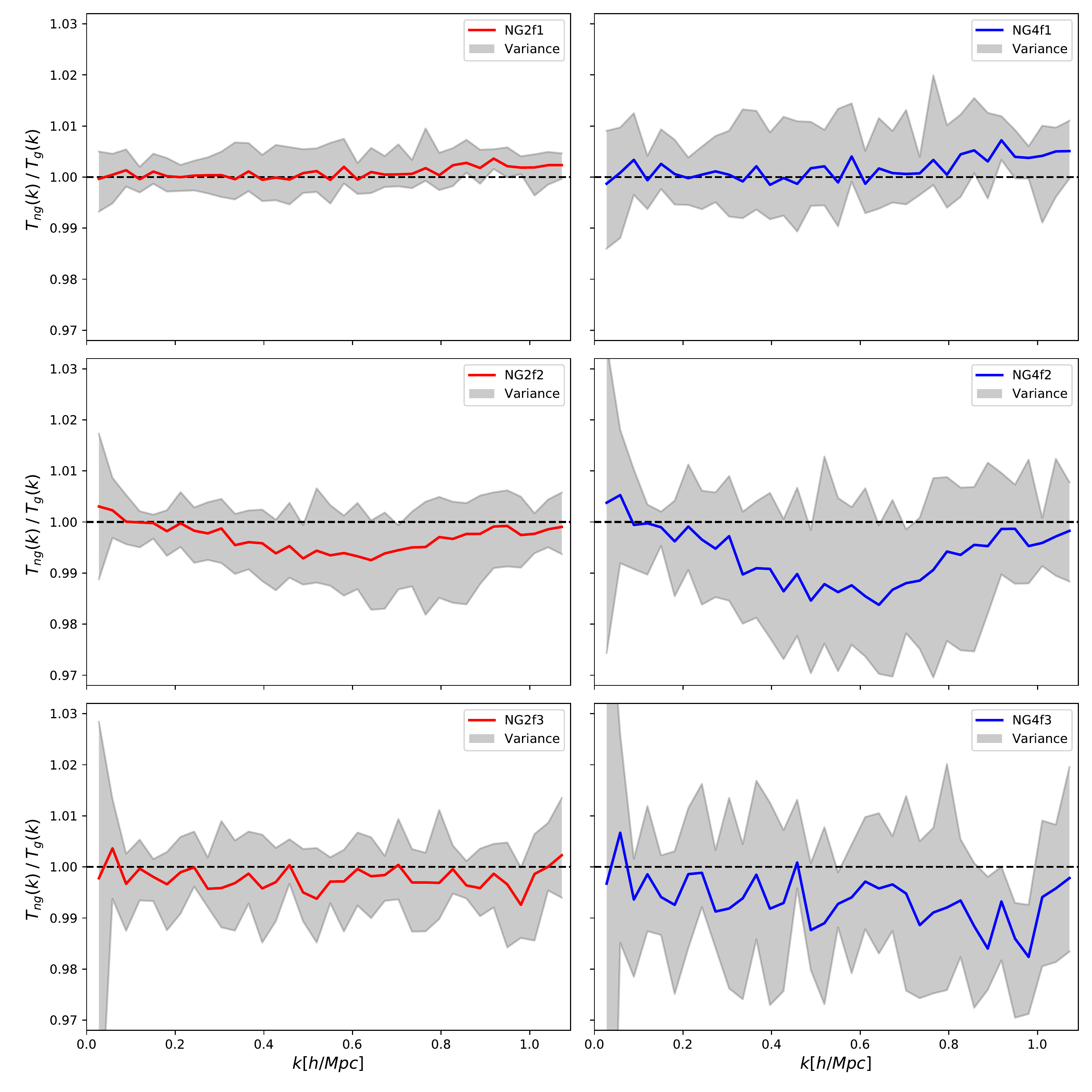}
    \caption{Non-Gaussian trispectra normalised with respect to the Gaussian trispectra, at $z=0$. The panels are as for Fig.~\ref{power_spec_realisations_z0}.}
    \label{Tk_realisations_z0}
\end{figure*}

\begin{figure*}
    \centering
    \includegraphics[width=0.8\linewidth]{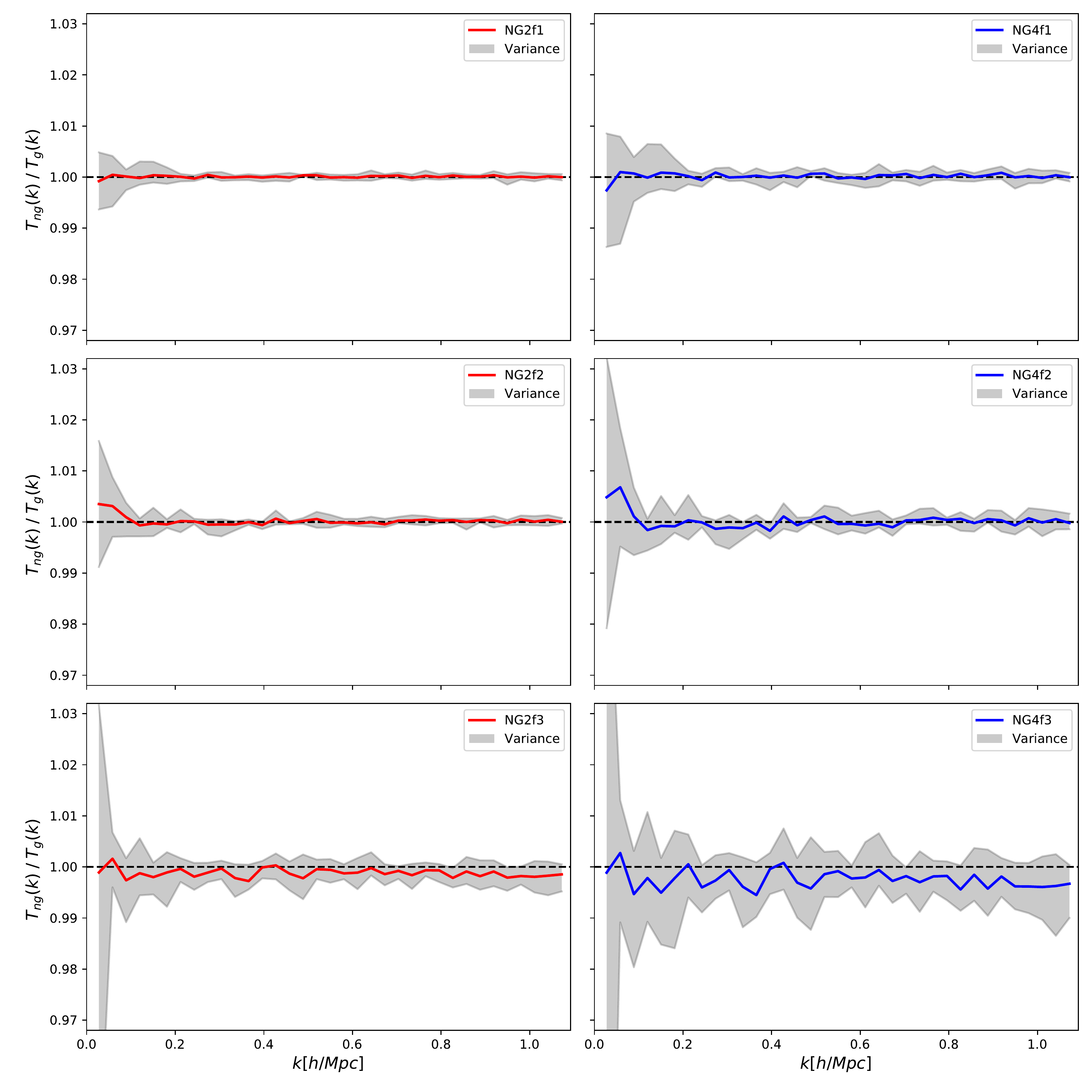}
    \caption{Non-Gaussian trispectra normalised with respect to the Gaussian trispectra, at $z=2.2$. The panels are as for Fig.~\ref{power_spec_realisations_z0}.}
    \label{Tk_realisations_z2.2}
\end{figure*}

In a similar manner to our analysis of the bispectra, we simplify the treatment of the trispectra by only considering ``square'' configurations of the four wavenumbers on which the trispectrum depends, i.e. $k_1=k_2=k_3=k_4$. This allows us to treat the trispectrum as a function of only a single value of $k$. We reiterate that although we use the ``square'' terminology, the $k$-space configurations are generally folded quadrilaterals, as discussed earlier. We place no restriction on the two additional degrees of freedom present in these configurations beyond the equal side lengths. It is worth remembering that the trispectrum is the Fourier space analogue of the \textit{connected} $4$-point correlation function, with the disconnected parts given by products of power spectra, corresponding to disconnected $2$-point correlation functions \citep{Verde2001}. Thus the trispectrum, even for our ``square'' configurations, is an independent statistical measure beyond the power spectrum.

The results for the averaged normalised trispectra for our various models, at $z=0$, are shown in Fig. \ref{Tk_realisations_z0}. It is interesting to note that the ``dip'' seen in the power spectrum and bispectrum at scales $0.2 \lesssim k \lesssim 0.8$ h/Mpc for the models with frequency f2 is also evident in the trispectrum. A note of caution must be struck regarding the vertical axis scale in Fig.~\ref{Tk_realisations_z0}, which is comparable to that of Fig.~\ref{power_spec_realisations_z0} for the power spectra, and therefore much reduced compared to that of Fig.~\ref{bk_realisations_z0} for the bispectra. Thus in absolute terms, the ``dip'' seen here is smaller than that seen in the bispectra. Again, this will be more carefully analysed in Section~\ref{sec:fitting_functions}.

In the case of the other frequencies there is very little obvious scale dependence, again as seen for the power spectrum and bispectrum. The f3 models show some indications of a suppression of the trispectrum across all accessible scales, with the smaller scales more suppressed in the NG4f3 model. It is also worth pointing out that the variance over realisations in the trispectra is larger than that of the power spectra, but lower than seen for the bispectra, being at most $2\%$ for the NG4f3 model.

The time evolution of the averaged normalised trispectra may be inferred from comparison with the results presented in Fig.~\ref{Tk_realisations_z2.2} corresponding to $z=2.2$. Again we see far less variance when compared to the $z=0$ results, and much less variance than the bispectrum at this same redshift, being more comparable with the power spectrum results. There is some indication in the NG4f3 model, however, that the suppression of the trispectrum is already evident, including the increase in this suppression at smaller scales. The variance again implies, however, that these results are not statistically significant.

\subsection{Fitting functions}
\label{sec:fitting_functions}


    \begin{figure*}
        \centering
        \includegraphics[width=0.8\linewidth]{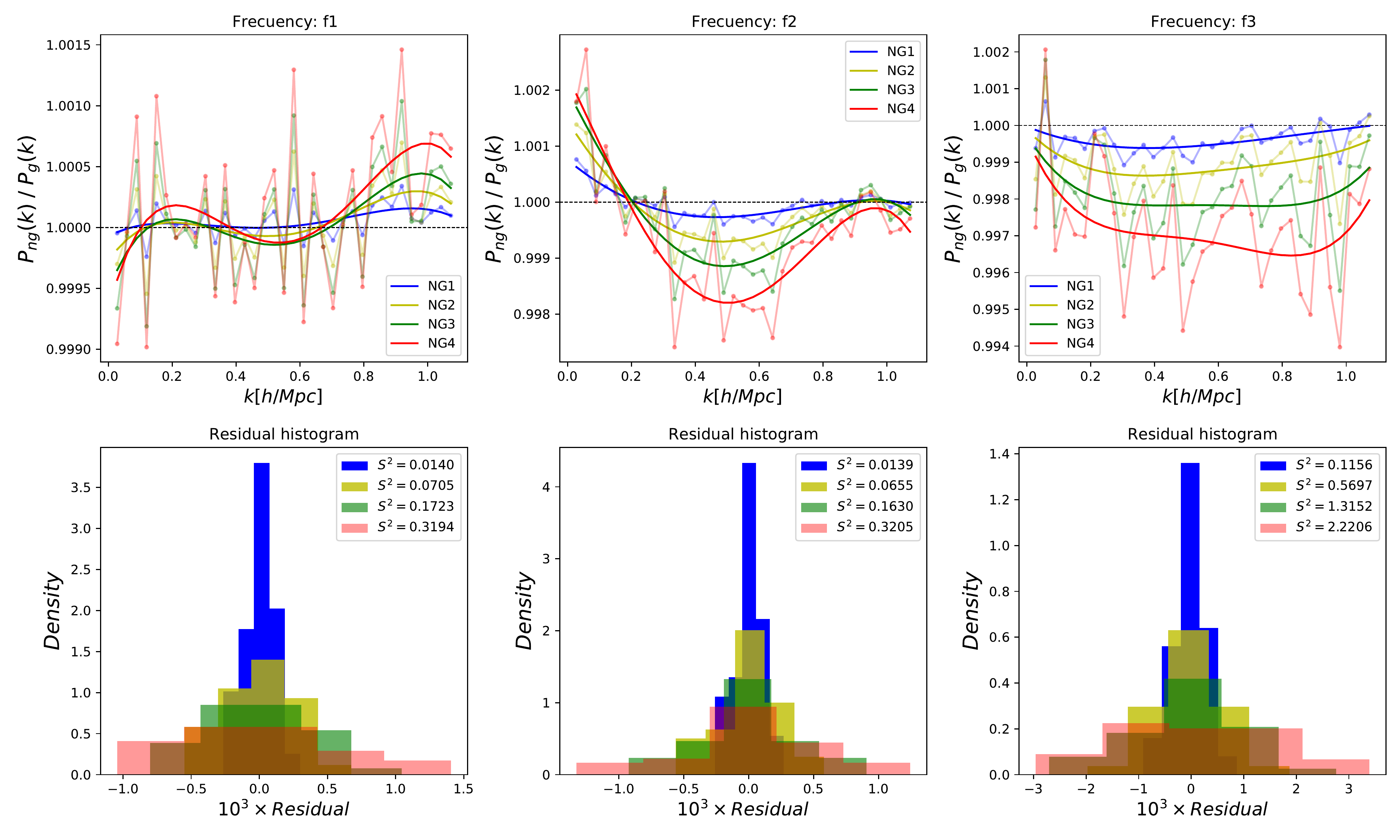}
        \caption{Fit lines for the averaged normalised power spectra at $z=0$. A 4th order polynomial fit is used for all frequencies (f1, left panel; f2, middle panel; f3, right panel). The variance $S^2$ of the residual histograms is also given (multiplied by $10^3$ to improve legibility of the plot).}
        \label{ps_fit}
    \end{figure*}

    \begin{figure*}
        \centering
        \includegraphics[width=0.8\linewidth]{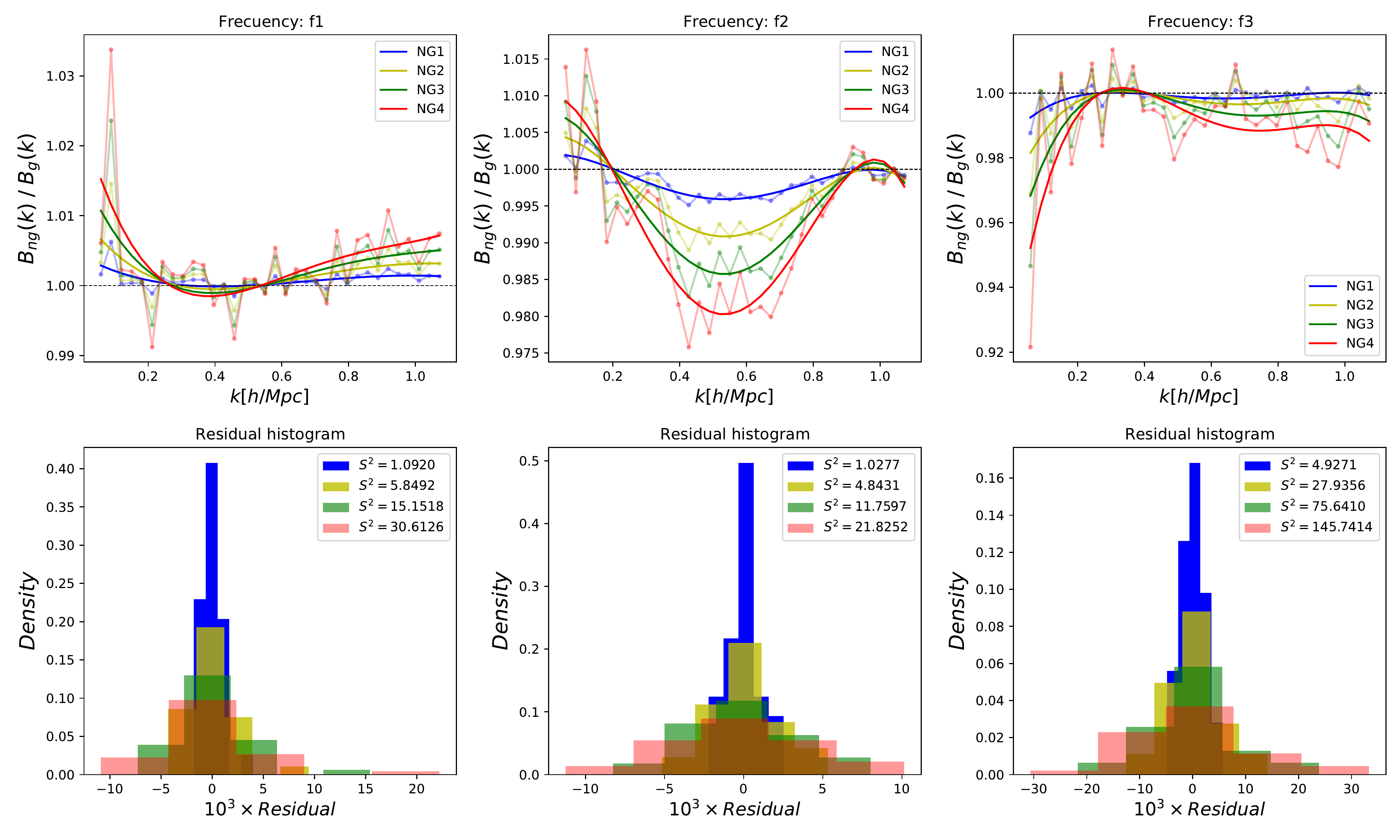}
        \caption{As for Fig.~\ref{ps_fit} with the fit lines and residuals for the averaged normalised
        bispectra at $z=0$.}
        \label{bis_fit}
    \end{figure*}
    
    \begin{figure*}
        \centering
        \includegraphics[width=0.8\linewidth]{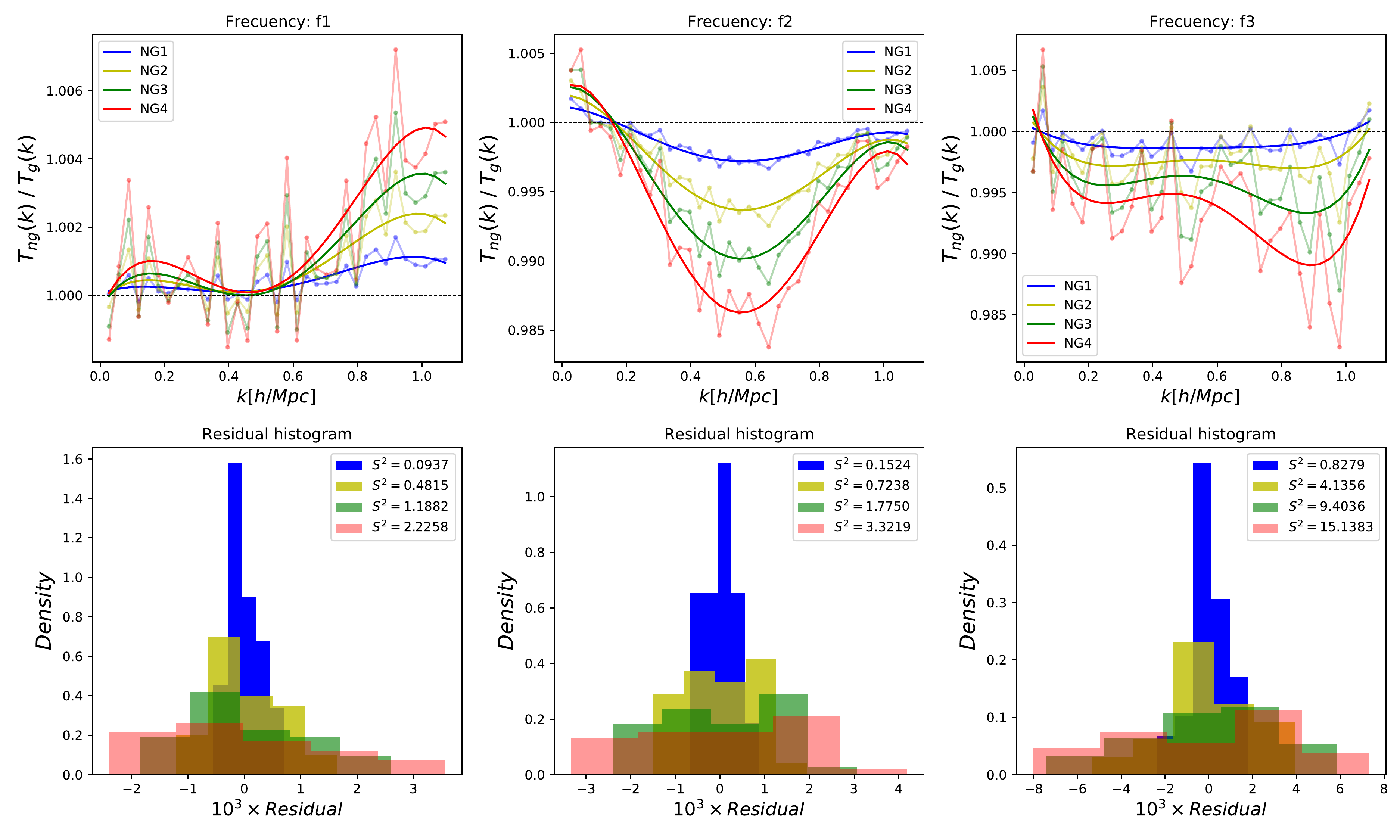}
        \caption{As for Fig.~\ref{ps_fit} with the fit lines and residuals for the averaged normalised trispectra at $z=0$.}
        \label{tris_fit}
    \end{figure*}

To further analyse the deviations from the Gaussian case in more detail, we will now determine best-fit lines for the averaged normalised polyspectra. This will allow us to determine the degree of deviation from the Gaussian case, characterise any possible systematic scale-dependence in $\Delta_{\mathcal{P}}(k)$, as well as the intrinsic properties of these quantities themselves for different values of $k$. 

We have found that a 4th order polynomial is the lowest degree polynomial to provide a reasonable fit for all frequencies. The goodness of the fits may be judged from the histograms of the residuals, which are plotted, along with the fits, for the power spectra in Fig.~\ref{ps_fit}, with the frequencies f1, f2 and f3 given in the left, middle and right panels. Similarly we show the fits for the bispectra in Fig.~\ref{bis_fit} and for the trispectra in \ref{tris_fit}. We discriminate between the levels of non-Gaussianity in each panel using coloured lines as indicated in the legend.

Comparing the results for frequency f1 across polyspectra (i.e. the left panels in Figs.~\ref{ps_fit}, \ref{bis_fit} and \ref{tris_fit}), we can see that the non-Gaussianity induces a systematic shift upwards at smaller scales in this case ($k \gtrsim 0.7$) with larger scales essentially unaffected. The magnitude of this effect depends on the polyspectrum considered: for the power spectrum it is extremely small, being at most $\sim 0.05\%$. For the bispectrum there is a very similar scale dependence, but with a stronger signal of $\sim 0.6\%$. The trispectrum exhibits, yet again, a similar scale dependence, with the deviation from Gaussianity at most $\sim 0.5\%$. In all cases the residual histograms are unimodal and approximately centred at zero, with some mild skewness arising for the higher levels of non-Gaussianity. It is also worth pointing out that the spread in the residuals increases with the level of non-Gaussianity, reflecting the increased variability in the averaged normalised polyspectra as the amplitude of the non-Gaussianity is increased.

Moving on to the frequency f2 (middle panel in the Figs.~\ref{ps_fit}, \ref{bis_fit} and \ref{tris_fit}) we see a different scale dependence in the deviation from Gaussianity, reflecting the ``dip'' feature referred to earlier. Again this feature appears for all polyspectra, with a maximum amplitude of $\sim 0.15\%$ in the power spectrum, $\sim 2\%$ for the bispectrum and $\sim 1.2\%$ in the trispectrum. The residual histograms for frequency f2 across polyspectra are again approximately centred at zero, with only slight skewness in some cases. The spread in the residuals is noticeably larger in the case of the bispectrum, indicating the increased variability of this statistic with increased non-Gaussianity.

Finally, in the case of frequency f3 (right panel of Figs.~\ref{ps_fit}, \ref{bis_fit} and \ref{tris_fit}) the long-wavelength deviations in the primordial PDF lead to deviations from the Gaussian model across all (accessible) scales, with a somewhat larger deviation at smaller scales. The form of the best-fit lines is very similar across all polyspectra, with the maximum amplitude of deviation from Gaussianity in the power spectrum being $\sim 0.4\%$, in the bispectrum $\sim 1.5\%$ and $\sim 1\%$ for the trispectrum. The residual histograms are mostly centred at zero with only mild skewness for the power spectra and bispectra, but there is more significant skewness of these histograms for the trispectra, perhaps indicating that the best-fit lines are overestimating the offset from the line of equality (where $T_{ng}/T_g = 1$) for the NG3 model, and underestimating for the NG2 and NG4 models. The spread in the residuals is also larger for the power spectra and trispectra compared to the other frequencies, and much larger for the bispectra. 

We also note that the residual spread, across all amplitudes of non-Gaussianity and all polyspectra, increases substantially for the frequency f3, compared with the other frequencies.

We should caution that the analysis of the residuals is affected by the low numbers of points that we are considering (we calculate the polyspectra using $35$ values in $k$-space, as stated earlier).

It is of interest to note that the most extreme case that we study (that of the NG4f3 model) does not lead to the largest deviations from Gaussianity in the $z=0$ polyspectra (at least up to the trispectrum). It is rather the NG2f3 model that exhibits the largest deviations, and we will verify that this model is apparently consistent with CMB constraints in the following section.

\subsection{CMB constraints}
\label{sec:CMBconstraints}

We will now verify that the models we have considered are consistent with constraints on the trispectrum derived from Planck observations of the CMB. We follow closely the analysis in \cite{PDFpaper}.

Firstly, however, we must relate our parameters for the asymptotic PDF given by equation~(\ref{asymptotic pdf}) with those of the full PDF given by equation~(\ref{eq:fullPDF}), for which the CMB constraints apply. We have found that a close match between the PDFs may be obtained by simply scaling up the values of $A^2$ given in Table~\ref{tab:A_f} when used in the full PDF. Thus we replace $A^2$ in equation~(\ref{eq:fullPDF}) with $\alpha A^2$. We find that the value of $\alpha$ required depends most sensitively on the frequency. For the f1 models we use a scaling of $\alpha = 33.64$, for the f2 models we use $\alpha = 11.56$ and for the f3 models we use $\alpha = 1.69$. For the most extreme model, NG4f3, we find the deviation of the asymptotic form from the full PDF (after applying the scaling) is $12\% \pm 11\%$, where the error quoted is one standard deviation around the mean. For all other models the deviation is smaller. The largest deviations occur in the tails of the distribution.

\begin{figure*}
  \centering
  \includegraphics[width=0.56\linewidth]{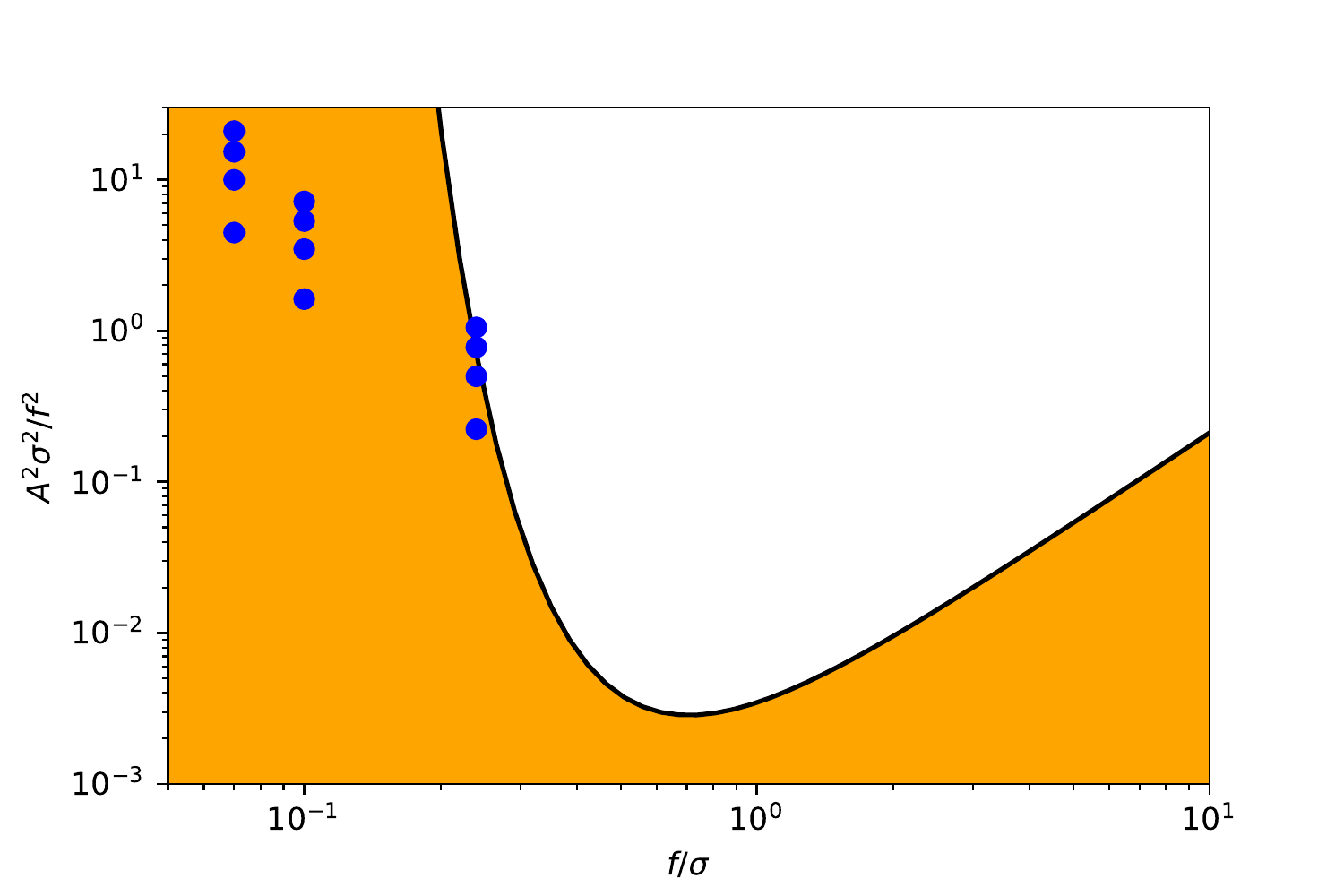}
  \caption{The permitted region of parameter space for the full PDF, as determined from the trispectrum constraint, is shown in orange. The excluded region is shown in white. The models we have considered are shown as blue dots.}
  \label{compare_CMB}
\end{figure*}

Using our scaling we can determine where our models lie in the permitted parameter space, when compared to Planck observations of the CMB. In \cite{PDFpaper} the following bound on the PDF parameters is obtained from considering the Planck constraint on the local primordial trispectrum shape $g_{\text{NL}}$:
\begin{equation}
    \beta < 2.1 \times 10^{-3},
\end{equation}
where
\begin{equation}
\beta \equiv A^2 \frac{\sigma_{\zeta}^4}{f_{\zeta}^4} \exp \left( -\frac{\sigma_{\zeta}^2}{2f_{\zeta}^2} \right)
\end{equation}
We have used $\sigma_{\zeta}^2 = 1$ for all of our models. The models we have considered are shown along with the permitted region of parameter space in Fig.~\ref{compare_CMB} (c.f. Fig. 8 in \citealp{PDFpaper}).

We can see that almost all of our models are well within the permitted (orange) region, albeit with large values of the $A^2$ parameter. There are two models that lie just within the excluded region (white): NG3f3 and NG4f3, the two most extreme cases (these have $\beta = 2.3 \times 10^{-3}$ and $3.1 \times 10^{-3}$ respectively). The frequency f2 models, however, all lie within the permitted region. These models have shown interesting scale-dependent deviations from Gaussianity in the late-time polyspectra, most significantly in the bispectrum (Fig.~\ref{bk_realisations_z0}).

It is important to keep in mind that this constraint is determined entirely from the trispectrum. If we instead consider the full PDF of temperature fluctuations in the CMB (as discussed with respect to a different NG model in \cite{Chen2018b}) it is very likely that our primordial non-Gaussian PDFs would be excluded. The large amplitudes that we have studied in this work, however, allow us to investigate the feasibility of detecting this type of NG in the late-time $n$-point statistics. We leave a more detailed confrontation with observations for future work.

\section{Discussion and conclusions}\label{sec:conc}

In this paper we have studied the impact on large scale structure statistics of a novel form of primordial non-Gaussianity, which is fully characterised by a modified probability density function (PDF) of the primordial curvature perturbations, where the non-Gaussianity manifests as an oscillatory modulation of an underlying Gaussian. We have studied the effect of this non-Gaussianity by generating samples from this modified PDF which are then used in the generation of pre-initial conditions for cosmological N-body simulations. The primordial density fluctuations drawn from our non-Gaussian distributions are convolved with a standard $\Lambda$CDM transfer function (in Fourier space) in order to generate the appropriate high-redshift density field. We consider $4$ levels of non-Gaussianity and $3$ different frequencies for the oscillatory non-Gaussian correction (our two most extreme models are marginally excluded by CMB constraints). Our pre-initial conditions are then used to determine initial particle positions and velocities (using 2LPT) for a small selection of models for cosmological N-body simulations using the RAMSES code. We use the L-PICOLA mock catalogue generation code to generate $5$ realisations per model, leading to a total of $65$ cosmological volumes.

We have confirmed, by comparing with the RAMSES results, that the non-Gaussian polyspectra normalised by the Gaussian (for the same realisation) are well represented by the L-PICOLA code.

Our study has focussed on analysing the low-$n$ correlation functions (specifically their Fourier space analogues: the power spectrum, bispectrum and trispectrum) to search for possible signatures of this type of primordial non-Gaussianity at low redshift.

We have found that, using the averaged normalised polyspectra, there are some scale-dependent deviations from the Gaussian model, which may be at the limit of detectability. Interestingly, the most significant signal is seen in the bispectrum for the models with frequency f2. For all other models, even the most extreme case NG4f3 (which is actually excluded by CMB constraints on the trispectrum) we see only sub-percent deviations from Gaussianity, well within the sample variance of differing non-linear realisations.

Thus our best-case scenario, given by the model NG4f2 (which lies well within the permitted region of parameter space) shows deviations from Gaussianity in the bispectrum at the level of $2\%$. The statistical significance of this deviation by considering sample variance is rather marginal, but may be detectable if sufficient precision can be obtained. It is also noteworthy that the signal present in the frequency f2 models is a scale-dependent suppression of the power spectrum, the symmetric bispectrum and the symmetric trispectrum. Consideration of non-symmetric configurations in $k$-space may well lead to further insights, and a more extensive exploration of the parameter space may uncover models whose late-time $n$-point correlation functions show more significant deviations of a similar form. Such an exploration may help to disentangle the precise relationship between the non-Gaussian frequency and the scale-dependence of the deviations from Gaussianity.

It is important to emphasise that our study suffers from a number of limitations which we intend to improve upon in future work. Firstly, the size of the cosmological volume and the spatial and particle (mass) resolution should be improved. In this work we have limited ourselves to a rather small box of $500$ Mpc$^3$ and $256^3$ particles. This would improve the accuracy of our statistical clustering measures, and may help reduce the variance over realisations that we have observed. We would also obviously benefit from using more realisations. Secondly, as suggested earlier, we intend to consider more general configurations of $k_i$ for the bispectrum and trispectrum. We have performed a preliminary analysis of the bispectrum using non-equilateral triangles, and we have not seen a significant difference from the results using equilateral triangles. Nevertheless, a more complete analysis demands the use of non-symmetric $n$-point functions.

Finally, in this work we have not studied alternative probes of the non-Gaussianity in the large scale structure, in particular the halo mass function. The enhanced probability for certain ranges of density peaks might be expected to lead to an oscillatory modulation of the halo mass function across a range of masses. Further work on this is currently in progress. There are also other possible probes of the primordial PDF itself beyond the polyspectra, such as Minkowski functionals or other topological measures.

While our results suggest this type of non-Gaussianity will be very challenging to detect, further study of the parameter space, using more observables, may uncover unexpected and novel tests of these models, helping to shine a light on the inflationary landscape.

\section*{Acknowledgements}

The authors thank the anonymous referee for helpful comments that improved the manuscript. The authors thank Gonzalo Palma and Spyros Sypsas for very useful comments. The authors acknowledge financial support from FONDECYT Regular No. 1181708. GP thanks the Postgrado en Astrofísica program of the Instituto de Física y Astronomía of the Universidad de Valparaíso for funding. GP also wishes to thank Hector Gil-Marín for helpful correspondence, and Daniela Palma for the hours of discussion on this project. 

\section*{Data Availability}

The data underlying this article will be shared on reasonable request to the corresponding author.



\bibliographystyle{mnras}
\bibliography{example} 








\bsp	
\label{lastpage}
\end{document}